\documentclass[showpacs,twocolumn,prl]{revtex4}
\usepackage{epsfig}
\usepackage{amsmath}
\usepackage{graphicx}
\usepackage{amssymb} 
\usepackage{graphics}
\usepackage{psfrag}

\begin{document}

\title{Measurement of Aharonov-Bohm oscillations in mesoscopic metallic rings in the presence of high-frequency electromagnetic fields}

\author{Josef-Stefan Wenzler, Pritiraj Mohanty}

\affiliation{
Department of Physics, Boston University, 590 Commonwealth Avenue, Boston, MA 02215}

\begin{abstract}
We report measurements of Aharonov-Bohm oscillations in normal metal rings in the presence of high frequency electromagnetic fields. The power dependence of the decoherence time scale $\tau_{\phi}(P)$ agrees well with the anticipated power law $\tau_{\phi}\propto P^{-1/5}$ when the field-induced decoherence rate $\tau_{ac}^{-1}$ is large compared to the intrinsic decoherence rate $\tau_{o}^{-1}$, measured in the absence of external fields. As theoretically expected, we observe a decline in field-induced decoherence when $\tau_{ac}^{-1}\leq\tau_{o}^{-1}$. The frequency dependence of $\tau_\phi$ shows a minimum in the oscillation amplitude at a characteristic frequency, $\omega_{ac} \simeq 1/\tau_{o}$, where $\tau_{o}$ is evaluated from the oscillation amplitude using the standard mesoscopic theory. Both the suppression in the oscillation amplitude and the concomitant change in conductivity allow a direct measurement of the intrinsic decoherence time scale. 

\end{abstract}


\maketitle

The decoherence time $\tau_\phi$, the time over which an electron retains its phase information, is of fundamental importance in condensed matter physics \cite{laurent}. Quantum interference effects such as weak localization (WL), universal conductance fluctuations (UCF) and Aharonov-Bohm (AB) oscillations in mesoscopic metal samples are characterized by the decoherence time, and are limited by decoherence mechanisms such as electron-electron interactions and magnetic impurity spins at temperatures below 1\hspace{0.5mm}K. Since intrinsic decoherence mechanisms and their low temperature dependence explore the ground state of the many-body metallic system \cite{mohanty-T,laurent}, precise determination of $\tau_\phi$ is fundamentally important.

In spite of its obvious importance, however, a direct measurement of $\tau_\phi$ in mesoscopic normal metal samples is yet to be reported. Instead, $\tau_\phi$ is obtained as a fit parameter, using standard models for quantum corrections to the electrical conductivity, such as WL, UCF and AB effect. There are several problems in determining $\tau_\phi$ in this way. First, a proper calculation of the decoherence rate due to the coupling to environmental degrees of freedom is usually not contained in a single self-consistent framework, except for weak localization ~\cite{golubev}. Second, the phenomenological introduction of suppression of quantum corrections with temperature (thermal averaging) or decoherence is usually done in an {\it ad hoc} manner by the inclusion of exponential and multiplicative suppression factors. Lastly, uncertainties in the input parameters, such as sample length, or in the definitions of quantities such as thermal diffusion length lead to a lack of precision in defining decoherence time~\cite{miliken,mohanty,pieper,chand} relevant to the experiments. Therefore, a direct measurement of $\tau_\phi$, independent of the specific model of quantum corrections, could help shed more light on decoherence mechanisms.

In their pioneering work~\cite{altshuler-SSC}, Altshuler, Aronov and Khmelnitsky have shown that electric fields of power $P$ can cause significant decoherence in metal wires as long as $\tau_{ac}^{-1}$, the decoherence rate due to the external high-frequency (hf) electric field, is large compared to the intrinsic decoherence rate $\tau_{o}^{-1}$, defined in the absence of the external electric field. In this limit, the total decoherence rate $\tau_{\phi}^{-1}=\tau_{o}^{-1}+\tau_{ac}^{-1}$ leads to $\tau_{\phi}\simeq\tau_{ac}$, where $\tau_{ac}\propto P^{-1/5}$ .  If $\tau_{o}^{-1}>\tau_{ac}^{-1}$,  a decrease in the field-induced decoherence efficiency is expected, and $\tau_{\phi}\simeq\tau_{o}$ should be recovered. The authors also suggest that maximum decoherence occurs for frequencies around $\omega\simeq\tau_{\phi}^{-1}$ when $\tau_{ac}\ll\tau_{o}$. Assuming decoherence mechanisms are the same in a ring geometry, a direct measurement of $\tau_{\phi}$ may be possible by studying the frequency dependence of the decoherence time $\tau_\phi (\omega)$.

In particular, for the frequency and power dependence of $\tau_\phi$, the study of AB oscillation amplitude is most suitable as small changes in input parameters can cause significant changes in $\tau_\phi$, due to the exponential dependence of the AB-oscillations on $\tau_\phi$. In comparison, weak localization and conductance fluctuation depend only weakly on $\tau_\phi$~\cite{mohanty}. Previous high frequency studies have mostly focused on weak localization effects~\cite{lindelof,wei,bykov,webbwkl}. In studies of the AB effect under high power, high frequency fields, Bartolo \textit{et al.} have observed a decline in the oscillation amplitude~\cite{bartolo}. In a different approach, Pieper and Price have attempted to observe strong decoherence by measuring AB oscillations with a high frequency current itself~\cite{pieper}.

The AB effect in phase coherent, normal metal rings of length $L$, threaded by a vector potential, $\vec{A}$, was first observed by Webb \textit{et al.}~\cite{webb-AB}. They measured small oscillations in the magnetoresistance of a mesoscopic gold ring with a period of h/e, the flux quantum. These oscillations arise from self-interference of single electrons~\cite{AB} passing through the ring that acquire a phase due to the vector potential, $(e/\hbar)\int \vec{A}\cdot d\vec{l}$.  The amplitude, $\Delta G_{AB}$, of the AB oscillations in the electrical conductivity  or resistance ($\Delta R_{AB}=-R^{2}\Delta G_{AB}$) is on the order of the quantum conductance $e^2/h$. It is reduced due to finite decoherence length $L_\phi=\sqrt{D\tau_\phi}$ and finite temperature $T$ so that
\begin{equation}
\Delta R_{AB}\simeq C\frac{e^{2}R^{2}}{h}\left(\frac{L_{T}}{L}\right)\sqrt{\frac{L_{\phi}}{L}}e^{-\frac{L}{L_{\phi}}},
\label{eq1}
\end{equation}
where $L_T=\sqrt{\hbar D/k_BT}$ is the thermal diffusion length, $D$ is the diffusion constant, $R$ is the resistance of the ring, and $C$ is a constant of order unity~\cite{washburn}. The factor $\sqrt{L_{\phi}/{L}}$ in Eq.~\ref{eq1} only applies when $L_{\phi}<L$. Hence, the equation above can be used to obtain $\tau_{\phi}$ from the AB oscillation amplitude, $\Delta R_{AB}$. Yet, multiple definitions used for $L$ and $L_{T}$~\cite{miliken,mohanty,pieper,chand}, as well as significant uncertainties on sample dimensions, constants and other input parameters limit the resolution of $\tau_{o}^{-1}$ to an order of magnitude. Even under ideal conditions for given definitions, $C=1$, and minimal uncertainties in input parameters (1\hspace{0.5mm}nm resolution in sample dimensions, and 1\% error on all other input parameters) $\tau_{o}^{-1}$ as obtained by Eq.~\ref{eq1} can still vary by more than 400~\hspace{0.5mm}MHz. This lack of precision causes problems for understanding decoherence mechanisms, and for numerical comparison between experiments and theory, which could be avoided with a direct measurement of $\tau_\phi$.

Here, in this Letter, we study frequency and power dependence of the field-induced coherence time, $\tau_{ac}$, in an attempt to directly identify the relevant time scale for decoherence. We find that the power dependence of $\tau_{\phi}$ (extracted from Eq.~\ref{eq1}) agrees well with the anticipated power law $\tau_{\phi}\sim P^{-1/5}$ when $\tau_{ac}^{-1}>\tau_{o}^{-1}$, and that the field-induced decoherence efficiency declines, when $\tau_{ac}^{-1}\leq\tau_{o}^{-1}$~\cite{altshuler-SSC}. We also observe strong decoherence at a specific frequency of $\omega_{ac} \simeq\tau_{o}^{-1}$ along with a concomitant dip in the resistivity at the same frequency. The predicted power dependencies of $\omega_{ac}$ was not observed. We suggest that the location of these minima in frequency can be used as a direct measurement of the intrinsic decoherence time $\tau_o$.

The two mesoscopic rings, data from which are reported here, are fabricated using standard e-beam lithography and thermal evaporation of $t=30\hspace{0.5mm}nm$ thick, ultra-pure (99.9999\%) silver (Ag) on a SiO$_{2}$ substrate. 
Both rings enclose an area of $A=0.8\hspace{0.5mm}\mu m^{2}$, with a circumference of $L=3.9\hspace{0.5mm}\hspace{0.5mm} \mu m$ and gate-to-gate separation of $g\approx1.6\mu m$. Other sample characteristics are summarized in Tab.~\ref{tab1}. Ring 1 was used for the power dependence and ring 2 for the frequency dependence data. Similar trends were observed in other ring samples studied in the experiment.

%---------------------------------------------------------
\begin{table}[t]\footnotesize
\caption{Relevant sample characteristics at $T=0.31\hspace{0.5mm} K$. (Errors for $\tau_{o}$ and $\tau_{o}^{-1}$ obtained from FFT background.)}
\begin{tabular*}{0.48\textwidth}{@{\extracolsep{\fill}}cccccccccc}
\hline\hline
& Width&R&$\Delta R_{o}$\footnote{without external field}&$D_{\mathrm{diff}}$&$\tau_{o}$&$L_{\phi}^{a}$&$\tau_{o}^{-1}$\\
&[nm] &[$\Omega$]&[$m\Omega$]&[$m^{2}/s$]&[ns]&[$\mu$m]&[MHz]\\\hline
Ring 1&48.0&28.40&1.82&0.009&2.86$\pm$0.13&5.16&350$\pm$16\\
Ring 2&51.1&11.15&0.51&0.022&2.01$\pm$0.14&6.69&498$\pm$33\\
\hline\hline
\end{tabular*}
\label{tab1}
\end{table}
%----------------------------------------------------------

\begin{figure}[ht]
	\includegraphics[width=.45\textwidth]{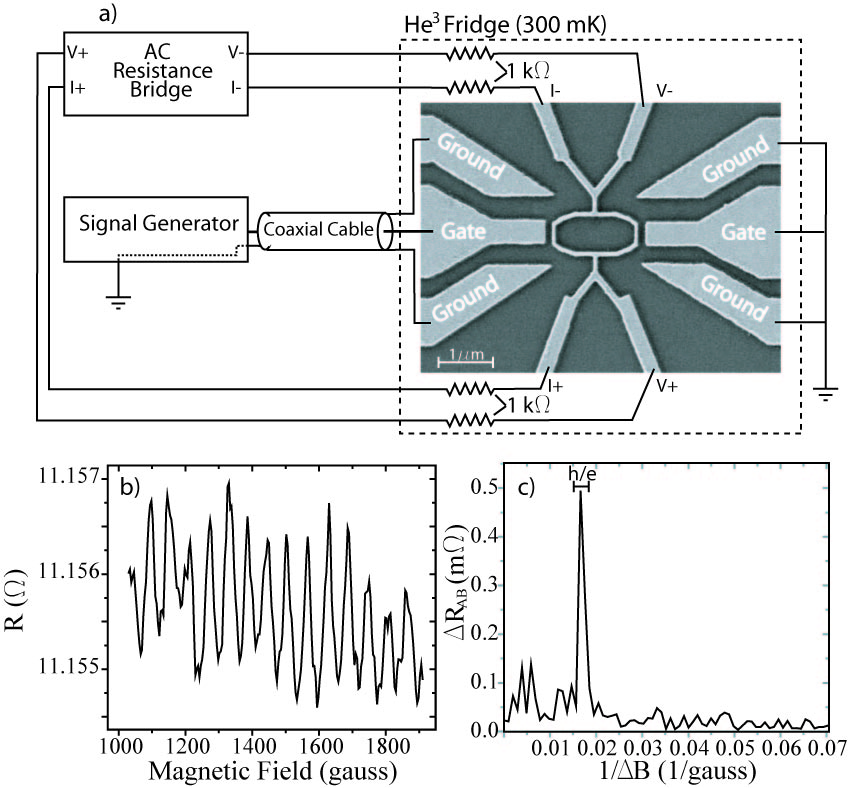} 
\caption{a) Schematic of the experimental setup. The resistance of the ring is measured with a standard 4-probe, ac resistance bridge technique, while the external field is applied within 100 nm of the gate via coaxial cables and coplanar waveguides. b) Unperturbed response of the magnetoresistance of ring 2, showing AB oscillations as expected. c) FFT of the AB oscillations is used to determine the amplitude $\Delta R_{AB}$ of the oscillations.
}
\label{pic1}
\end{figure}

The experimental setup is depicted in Fig.~\ref{pic1}a). The sample is mounted on a specially-designed
sample stage, attached to the mixing chamber of a $\mathrm{He}^{3}$ cryostat at $300\hspace{0.5mm}$mK. Resistance is measured, via low-frequency twisted pairs with 1\hspace{0.5mm}k$\Omega$ resistors, at $15.9\hspace{0.5mm}$Hz using a standard four-probe, ac resistance bridge technique, while a hf electromagnetic field (100\hspace{0.5mm}MHz--12.4\hspace{0.5mm}GHz) is applied in-plane to perturb the ring locally via the left gate. The hf field is guided within $d\simeq 100\hspace{0.5mm}nm$ of the ring utilizing $50\hspace{0.5mm}\Omega$-impedance coaxial cables and $\sim77\hspace{0.5mm}\Omega$-impedance micro-fabricated coplanar waveguides, as measured by time domain reflectometry.  
%----------------------------------------------------
%-------------Figure 0--------------------------------
%------------------------------------------------------
%------------------------------------------------------
%------------------------------------------------------
%------------------------------------------------------

As shown in Fig.~\ref{pic1}b), a typical unperturbed magnetoresistance measurement exhibits AB oscillations with the expected periodicity of $B = \frac{h}{eA}\simeq 5.2\hspace{0.5mm}$mT. The AB oscillation amplitude $\Delta R_{AB}$ is obtained from the Fast Fourier Transform (FFT) of the magnetoresistance, as shown in Fig.~\ref{pic1}c). The corresponding unperturbed decoherence time and length scales, as determined by Eq.~\ref{eq1}, are given in Tab.~\ref{tab1}.

\begin{figure}[ht]
	\includegraphics[width=.45\textwidth]{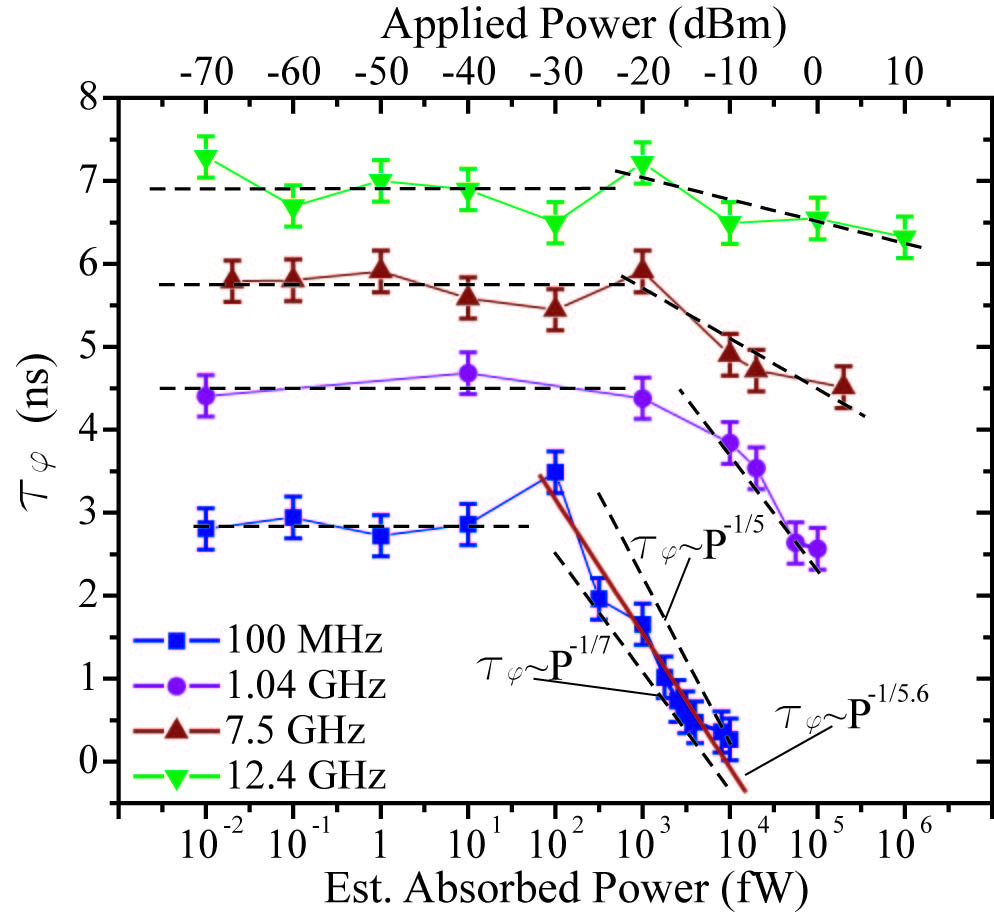}
\caption{(Color online) $\tau_{\phi}$ as a function of power absorbed by ring 1. The red solid line is a fitted power law $\tau_{\phi}\propto P^{-1/(5.6\pm0.5)}$, the dotted lines serve as guide to the eye. All plots have been adjusted for lossy cables and wire-bonds. Estimated absorbed power values are obtained from detailed analysis and measurement of transmission coefficients between the ring and gates. (1.04\hspace{0.5mm}GHz, 7.5\hspace{0.5mm}GHz  and 12.4\hspace{0.5mm}GHz plots are offset for clarification by 2.3\hspace{0.5mm}ns, 4\hspace{0.5mm}ns, and 6\hspace{0.5mm}ns respectively.)}  
\label{pic2}
\end{figure}

According to Ref.~\cite{altshuler-SSC}, $\tau_{ac}$ depends on the external field frequency and power. For the rings studied here, the condition $\tau_{ac}^{-1}\geq\tau_{o}^{-1}$ is satisfied at 1\hspace{0.5mm}GHz for applied powers above $P=-30\hspace{0.5mm}dBm$, at 7.5\hspace{0.5mm}GHz for applied powers above $P=-14\hspace{0.5mm}dBm$ , and  at 12.4\hspace{0.5mm}GHz for applied powers above $P=-9\hspace{0.5mm}dBm$. Therefore, as a function of power, demonstrated in Fig.~\ref{pic2}, we expect $\tau_{\phi}(P)\propto P^{-1/5}$ for 100\hspace{0.5mm}MHz, whereas for frequencies above 1\hspace{0.5mm}GHz a decline in field-induced decoherence efficiency is expected. 

This is indeed the case as illustrated in Fig.~\ref{pic2}. Each point in Fig.~\ref{pic2} represents $\tau_{\phi}$ obtained from the FFT of a 3-hour magnetoresistance sweep. The error bars for all figures are obtained from the corresponding FFT background. For the 100\hspace{0.5mm}MHz data, where $\tau_{ac}^{-1}\geq\tau_{o}^{-1}$, the red solid line represents a fitted power law $\tau_{\phi}\propto P^{-1/(5.61\pm0.5)}$. The dotted lines for the other frequencies in Fig.~\ref{pic2} (above $1\hspace{0.5mm}$GHz), where $\tau_{ac}^{-1}\leq\tau_{o}^{-1}$, represent guides to the eye to point out the anticipated decline in field-induced decoherence efficiency.

%---------------------------------------------------
%---------------Figure 2-------------------------
%--------------------------------------------------

%---------------------------------------------------
%---------------Figure 2-------------------------
%--------------------------------------------------

\begin{figure}[ht] 
	\includegraphics[width=.45\textwidth]{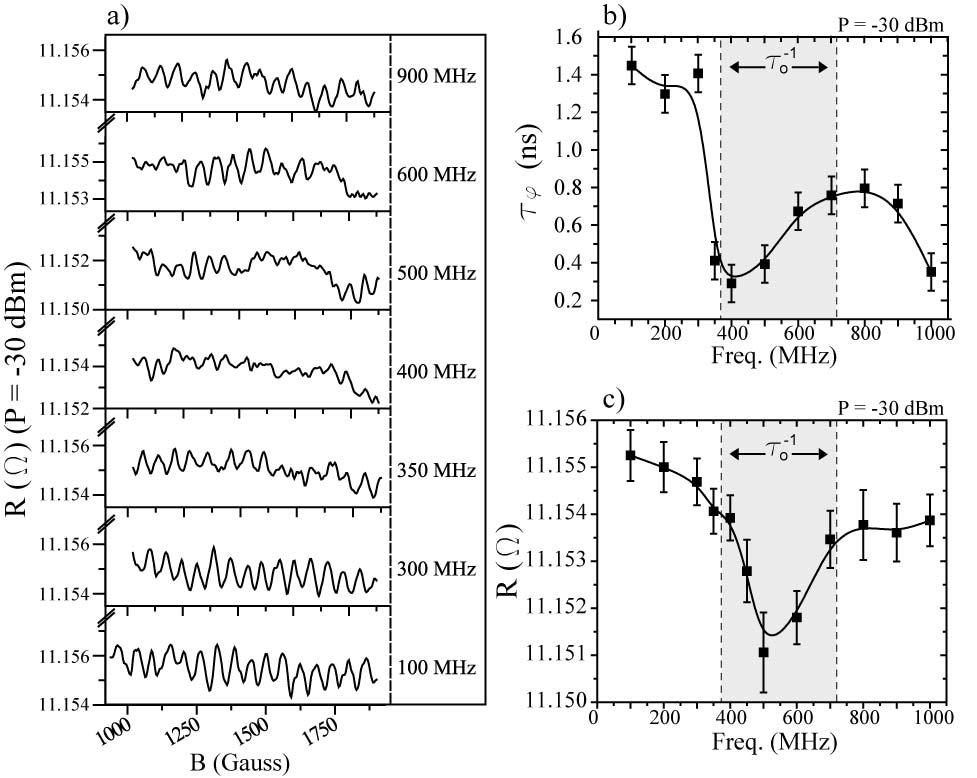}
	\caption{The lines connecting the data points in b) \& c) serve as guide to the eye and the shaded regions designate the range of values for $\tau_{o}^{-1}$ under ideal conditions as described in the text. Data was obtained measuring ring 2. a) AB oscillations for different pertubation frequencies at $P=-30\hspace{0.5mm}$dBm. Strong oscillations at 100\hspace{0.5mm}MHz disappear almost completely at 400\hspace{0.5mm}MHz and reemerge at 900\hspace{0.5mm}MHz. b) $\tau_{\phi}(\omega)$ extracted from Eq.~\ref{eq1} is depicted. At $\omega_{ac}\simeq\tau_{o}^{-1}$ a dip is observed. c) $R(\omega)$ as a function of perturbation frequency is shown. At $\omega_{ac}\simeq\tau_{o}^{-1}$, $R(\omega)$ has a minimum.}
\label{pic3}
\end{figure}

Fig.~\ref{pic3}a) shows the magnetoresistance in the presence of hf fields for various frequencies at an applied power of -\hspace{0.5mm}30\hspace{0.5mm}dBm. Following the sweeps from 100\hspace{0.5mm}MHz to 900\hspace{0.5mm}MHz, we observe two trends. First, there is a minimum in $\tau_{\phi}$ around 500\hspace{0.5mm}MHz, well within the range of theoretical predicted values for $\tau_{o}^{-1}$ under ideal conditions as described above. Second, the overall resistance of the ring $R(\omega)$ reflects this behavior in that it has a dip around $\tau_{o}^{-1}$ as well. This effect is about 4 times bigger than the AB amplitude. These two trends are singled out and depicted in Fig.~\ref{pic3}b)~and~c) respectively. 

\begin{figure}[ht] 
	\includegraphics[width=.45\textwidth]{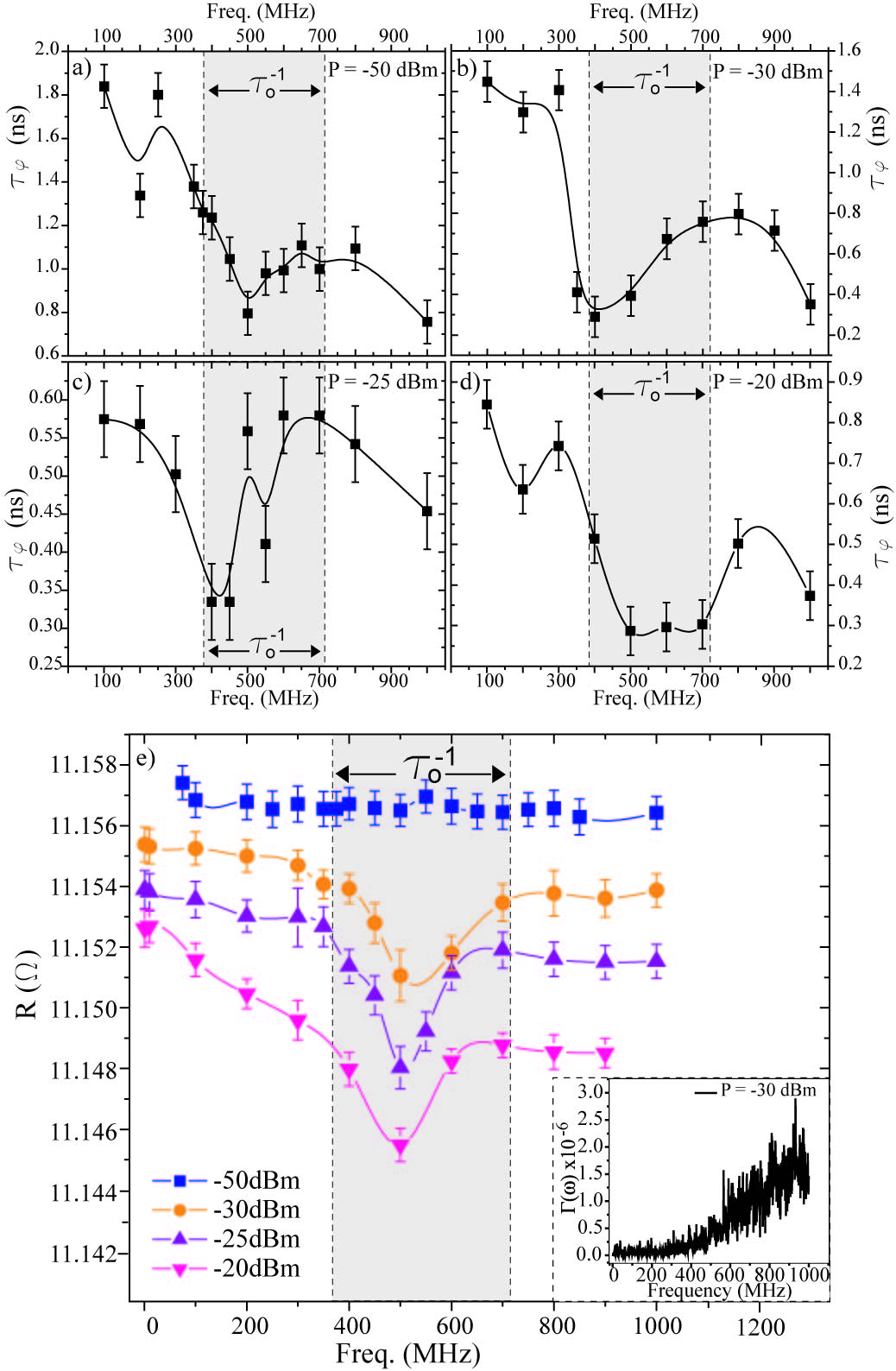} 
	\caption{(Color online) The lines through the data points serve as guide to the eye and the shaded regions demonstrate the uncertainty in $\tau_{o}^{-1}$ under ideal conditions as described in the text. a) $\tau_{\phi}(\omega)$ for ring 2 is depicted as for applied powers of -50\hspace{0.5mm}dBm. b) Shows $\tau_{\phi}(\omega)$ at -30\hspace{0.5mm}dBm. A sharp drop in $\tau_{\phi}$ is evident at $\omega_{ac}\simeq\tau_{o}^{-1}$. c) Depicts $\tau_{\phi}(\omega)$ at -25\hspace{0.5mm}dBm. d) At -20\hspace{0.5mm}dBm the system is strongly perturbed but again exhibits a minimum at $\omega_{ac}\simeq\tau_{o}^{-1}$. e) Composite graph of $R(\omega)$ of the ring for several different powers (-50\hspace{0.5mm}dBm, -25\hspace{0.5mm}dBm, and -20\hspace{0.5mm}dBm  data are offset for clarification by 1\hspace{0.5mm}m$\Omega$, -1\hspace{0.5mm}m$\Omega$, and -2\hspace{0.5mm}m$\Omega$ respectively). The inset demonstrates the frequency dependence of gate to gate transmission coefficient, $\Gamma(\omega)$. All graphs show lower values of $\tau_{\phi}(\omega)$ for higher frequencies which can be attributed to the monotonic increase in $\Gamma(\omega)$.
}
\label{pic4}
\end{figure}

The data sets indicate that both minima are due to more efficient decoherence at a characteristic frequency of $\omega_{ac}\simeq\tau_{o}^{-1}$. Maximum decoherence at $\omega_{ac}\simeq\tau_{o}^{-1}$ for various powers is depicted in Fig.~\ref{pic4}. The precision in the location of the minima suggests that it can be used for direct measurement of the intrinsic decoherence time scale. While the determination of $\tau_{o}^{-1}$ from Eq.~\ref{eq1} is rather imprecise ($\pm 170\hspace{0.5mm}$MHz) even under ideal conditions, it can be measured directly within $\pm25$\hspace{0.5mm}MHz as can be seen in Fig.~\ref{pic4}e).

Although a non-monotonic dependence like the minimum in $R(\omega)$ could potentially arise due to electron-electron interaction, corrections arising from diffusion diagrams are not expected to change at high frequencies. In any case, there are no existing calculations to enable a direct numerical comparison.
%---------------------------------------------------
%---------------Figure 3-------------------------
%------------------------------------------------------
%---------------------------------------------------
%---------------Figure 3-------------------------
%------------------------------------------------------
%---------------------------------------------------
%---------------Figure 4-------------------------
%--------------------------------------------------
%---------------------------------------------------
%---------------Figure 4-------------------------
%--------------------------------------------------

In order to avoid electron heating, observed here for low-frequency measurement powers of above $20~pW$, the power level is kept at $P_{dc}\simeq 10\hspace{0.5mm}pW$. We have also performed detailed measurement and analysis following Refs.~\cite{aleiner,kanskar,liu2,wei}, and we find that the minimum cooling power $P_{diff}$, provided by the out-diffusion of hot electrons into cooler pads, is larger than the maximum microwave heating power $P_{ac}$ due to the power absorbed by the electrons. 

The power absorbed by the ring due to the hf signals arises from the open-ended coplanar waveguide, substrate surface waves, and capacitive coupling between the gates with the ring in the middle~\cite{beilenhoff,wood,james}. Rigorous analysis shows that the contributions from the first two mechanisms are small ($\sim10^{-14}W$), compared to the contribution due to capacitive coupling. Therefore, to estimate the power absorbed by the ring we have measured the frequency dependent gate to gate transmission coefficients, $\Gamma(\omega)$, of hf signals released from the top of the He$^{3}$-refrigerator for several identical rings (see inset Fig.~\ref{pic4} e)). Using the maximum transmission coefficient, $\Gamma_{max}=7\cdot10^{-5}$ and the maximum applied voltage waves of $V_{max}=22\cdot10^{-3}V$, we estimated the upper limit of the power absorbed by the ring to be $\left[P_{ac}\right]_{max}=(V_{max}^{2}\Gamma_{max}^{2}L^{2})/(g^{2}R)\simeq2\cdot10^{-12}W$. While $\Gamma(\omega)$ monotonically increases with $\omega$ and thus explains the overall decline in $\tau_\phi$ in Fig.~\ref{pic3}~b) for higher frequencies, it does not explain the non-monotonic behavior at $\omega\simeq\tau_{o}^{-1}$.

The cooling power due to the out-diffusion of hot electrons into cooler current leads depends on the temperature gradient between the hot and cold electrons. Assuming the smallest significant temperature gradient of $\Delta T_{min} = 50\hspace{0.5mm}mK$, the minimum cooling power can be calculated~\cite{aleiner}:
$\left[P_{diff}\right]_{min}=\left(2\pi k_{B}T/e\right)^{2} (T\Delta T_{min}/R)\simeq 4\times10^{-11}\hspace{1mm}\mathrm{W}$.
Since $\left[P_{ac}\right]_{max}<\left[P_{diff}\right]_{min}$ joule heating can be neglected. 

However, $P_{ac}$ is large enough to cause decoherence. According to Ref.~\cite{wei,aleiner}, the electric field amplitude $E_{ac}$ required to cause strong decoherence for hf signals in the optimal frequency range of $\omega\tau_{\phi}\simeq1$ can be estimated as $eE_{ac}L_{\phi}\sim\hbar/\tau_{\phi}$. Therefore, the minimum power required to cause strong decoherence is:
$P_{\phi}=(E_{ac}L)^{2}/R\approx 3\times10^{-15}\hspace{2mm}\mathrm{W}$. This is in good agreement with our data, as strong decoherence is observed for applied powers from -20\hspace{0.5mm}dBm to -30\hspace{0.5mm}dBm, representing a power range experienced by the ring between $0.13\hspace{0.5mm}pW$ to $1.3\hspace{0.5mm}pW$ according to $\left[P_{ac}\right]=(V^{2}\Gamma^{2}L^{2})/(g^{2}R)$, but only weak decoherence is found for -50\hspace{0.5mm}dBm representing an estimated maximum absorbed power of $1.3\hspace{0.5mm}fW$ (see Fig.~\ref{pic4}e).

The correlation between the unique dependence of both $\tau_{\phi}(\omega)$ and the overall resistance  $R(\omega)$ can not be explained by the fact that $\tau_{\phi}$ depends on $R$, as the change in $R$ of ($\sim 4\hspace{0.5mm}m\Omega$) is too small to cause a significant change in $\tau_{\phi}$. Furthermore, non-monotonic dependencies, as seen in Fig.~\ref{pic4}a)-e), are not expected from joule heating, the standard electron interaction correction, $\Delta R_{ee}\sim\frac{e^{2}L_{T}R^{2}}{h L}$~\cite{laurent}, or the photovoltaic effect in mesoscopic metal rings~\cite{bartolo}. 

In conclusion, we have experimentally measured the dependence of Aharonov-Bohm oscillation amplitude in mesoscopic rings on external electromagnetic fields, applied controllably through a local gate. We find that for $\tau_{ac}^{-1}>\tau_{o}^{-1}$ the power dependence of coherence time agrees fairly well with $\tau_{\phi}(P)\propto P^{-1/5}$, while for higher frequencies when $\tau_{ac}^{-1}<\tau_{o}^{-1}$ the expected decline in decoherence efficiency is observed. At the characteristic frequency $\omega_{ac}\simeq1/\tau_{o}$ we observe strong decoherence and a corresponding dip in the resistivity. We suggest that the location of these minima can be used to determine $\tau_{o}$ directly.

This work was supported by the NSF (CCF-0432089).

\end{document}